\documentclass{svproc}
\usepackage{url}

\usepackage{cite}
\usepackage{amsmath,amssymb,amsfonts}
\usepackage{algorithmic}
\usepackage{graphicx}
\usepackage{textcomp}
\usepackage{xcolor}
\usepackage{hyperref}
\makeatletter
\let\orig@unskip\unskip
\renewcommand{\unskip}{\ifvmode\else\orig@unskip\fi}
\makeatother
\usepackage{booktabs}
\usepackage{tikz}
\usepackage{pgfplots}
\usepackage{placeins}
\usepackage{eso-pic}
\usepackage[stretch=10,shrink=10]{microtype}
\pgfplotsset{compat=1.18}
\usetikzlibrary{patterns, shapes.geometric, arrows.meta, positioning}
\begin{document}
	\setlength{\abovedisplayskip}{4pt}
	\setlength{\belowdisplayskip}{4pt}
	\mainmatter
	\title{The Memory Wall of Green Software: Empirical Energy Evaluation of Memento Design Pattern}
	\titlerunning{The Memory Wall of Green Software}
	\author{Imane JRIRI\inst{1} \and Tarik HOUICHIME\inst{2} \and Younes EL AMRANI\inst{1}}
	\authorrunning{I. Jriri et al.}
	\institute{\inst{1}LRIT, Faculty of Science, Mohammed V University In Rabat, Rabat, 10112, Morocco\\
		\inst{2}Meridian Team, LyRICA Laboratory, School of Information Sciences, Rabat 10100, Morocco\\
		\email{imane\_jriri@um5.ac.ma, thouichime@esi.ac.ma, y.elamrani@um5r.ac.ma}}
	\maketitle
	\AddToShipoutPictureBG*{%
		\AtPageUpperLeft{\raisebox{-1.1cm}{\hspace*{0.1\paperwidth}%
				\parbox{0.8\paperwidth}{\centering\scriptsize\itshape
					This is the authors' original preprint version of a paper accepted for
					publication at the International Conference on Advanced Materials for
					Sustainable Energy and Engineering (ICAMSEE~2026), Ifrane, Morocco,
					May 19--21, 2026. The final version will be published in the conference
					proceedings (indexed by Scopus).}}}%
		\AtPageLowerLeft{\raisebox{1.0cm}{\hspace*{0.1\paperwidth}%
				\parbox{0.8\paperwidth}{\centering\scriptsize\itshape
					This is the preprint version of a paper accepted for publication at
					ICAMSEE~2026. The final version will appear in the official conference
					proceedings (indexed by Scopus).}}}%
	}
	\begin{abstract}
		As Green Software Engineering matures, energy efficiency has transitioned into a mission-critical non-functional requirement. While software design patterns ensure structural integrity, their inherent abstraction layers impose an implicit {``metabolic cost''} that often remains obscured during the design phase. This paper empirically investigates the energy dynamics of the Memento design pattern, contrasting a direct, unabstracted baseline against {Classic full-snapshot} and {Differential delta-encoding} strategies. Leveraging the {RAPL} interface for high-fidelity hardware telemetry, we quantify energy dissipation across state volumes scaling from 10~MB to 200~MB. Our empirical results expose a critical architectural trade-off: the Differential strategy minimizes memory traffic, yielding a maximum {energy reduction of 65.8\%} for mid-scale states, but collides with a catastrophic {``memory wall''} at 200~MB. At this saturation point, algorithmic optimizations are completely neutralized by severe {GC} thrashing and non-linear power spikes. We synthesize these findings into evidence-based heuristics, providing architects with a robust framework to reconcile structural design quality with sustainable Green IT imperatives.
	\end{abstract}
	\section{Introduction}
	Energy efficiency has evolved from a peripheral operational metric to a primary design constraint in sustainable computing. While hardware-level interventions like \textit{Dynamic Voltage and Frequency Scaling} (DVFS) effectively mitigate peak power consumption, empirical literature demonstrates that software architectural decisions establish the baseline energy footprint---often eclipsing the marginal gains of hardware optimizations \cite{Sahin2012, Pereira2017}. Fundamentally, software dictates the metabolic rate of underlying resources.
	Although design patterns provide essential scaffolding for software modularity \cite{Gamma1995}, these abstractions are rarely energy-neutral. In managed runtimes, layers of indirection incur a measurable runtime overhead that is systematically underrepresented in design-time decision-making \cite{Georgiou2023, Bree2025}. This study isolates the \textit{Memento} pattern, the ubiquitous paradigm for state recovery. Production implementations typically adopt either a \textit{Classic} full-state snapshot approach or a \textit{Differential} variant leveraging delta encoding. While the Differential approach is favored for its minimal memory footprint, its true thermodynamic implications remain ambiguous. Specifically, does the CPU overhead of delta computation justify reduced memory traffic, or does it inadvertently trigger catastrophic garbage collection (GC) thrashing \cite{Melfi2023}?
	To resolve this ambiguity, we conduct a rigorous empirical evaluation using the Intel \textit{Running Average Power Limit} (RAPL) interface \cite{Khan2018}. Our experimental analysis yields four critical insights:
	\begin{itemize}
		\item \textbf{The Sustainability Sweet Spot (50--100\,MB):} Within this bounded scale, the Differential strategy curtails DRAM activation, yielding a peak energy reduction of \textbf{65.8\%} relative to the Classic approach.
		\item \textbf{The Memory Wall (150--200\,MB):} We document a severe efficiency inversion. Beyond this threshold, Generation 2 GC thrashing escalates by \textbf{2,300\%}, causing the Differential pattern to consume \textbf{25.9\% more energy} than the na\"ive Classic implementation.
		\item \textbf{Metabolic Dominance:} At high object volumes, runtime dynamics (specifically deep heap management) dominate the energy budget.
		\item \textbf{Architectural Heuristics:} We formalize these findings into a state-aware decision matrix to guide sustainable software architecture.
	\end{itemize}
	\begin{table}[ht]
		\centering
		\caption{Symbols and definitions utilized in the formalized energy models.}
		\scriptsize
		\setlength{\tabcolsep}{4pt}
		\begin{tabular}{@{}lp{3.8cm}lp{3.8cm}@{}}
			\toprule
			\textbf{Sym.} & \textbf{Description} & \textbf{Sym.} & \textbf{Description} \\
			\midrule
			$E_{Total}$ & Total cumulative energy (J) & $P_{Core}$ & Mean CPU active power (W) \\
			$P_{DRAM}$ & Mean RAM active power (W) & $S_{cl}, S_{diff}$ & Classic \& Differential models \\
			$\Omega$ & Memory footprint (MB) & $\phi_{GC}$ & GC frequency (Hz) \\
			$\kappa_{IO}$ & Serialization cost (J/MB) & $\gamma_{CPU}$ & CPU compute cost (J/cycle) \\
			$\delta_{DRAM}$ & RAM activation cost (J/MB) & $\Omega_{GC}$ & GC energy overhead (J) \\
			\bottomrule
		\end{tabular}
	\end{table}
	\FloatBarrier
	\section{Related Work}
	Green Software Engineering has rapidly matured from qualitative heuristics to rigorous empirical methodologies grounded in hardware telemetry.
	\textbf{Energy Telemetry Fidelity:} High-resolution energy analysis necessitates robust measurement mechanisms. Intel's RAPL interface is recognized as the industry standard for granular, non-intrusive energy telemetry \cite{Khan2018, Phung2018}. Software-level estimation models inherently introduce an ``observer effect'', perturbing the baseline profile. To circumvent this bias, our methodology relies exclusively on hardware-level RAPL data acquisition \cite{Ferreira2023, Noureddine2022}.
	\textbf{Architectural Influence:} Software architecture acts as the fundamental blueprint dictating system energy demand \cite{Procaccianti2016, Manotas2016}. While structural patterns have been extensively profiled \cite{Bree2025, Callau2023}, state-intensive behavioral patterns remain critically underexamined \cite{Georgiou2023}. Evaluating \textit{Memento} requires balancing the computational overhead of state traversal against the memory bandwidth demanded by persistent storage.
	\textbf{The Managed Runtime Paradox:} The metabolic footprint of architectural abstractions must be evaluated alongside their execution environment. In managed runtimes (.NET, JVM), energy variance is frequently dictated by heap allocation heuristics rather than pure CPU throughput \cite{Pereira2017, Oliveira2021}. Suboptimal memory utilization triggers disproportionate energy spikes \cite{Pinto2016, Garcia2024}, validating the ``memory wall'' hypothesis \cite{Hiziroglu2025}, suggesting algorithmic optimizations are negated by deep heap maintenance costs \cite{Hindle2012}.
	\section{Experimental Methodology}
	To ensure empirical reproducibility and isolate software-induced energy variance, we engineered a highly controlled experimental protocol synthesizing hardware telemetry with rigorous statistical controls.
	\begin{figure}[!ht]
		\centering
		\begin{minipage}[t]{0.48\textwidth}
			\vspace{0pt}
			\centering
			\resizebox{\linewidth}{!}{%
				\begin{tikzpicture}[
					node distance=0.2cm,
					layer/.style={rectangle, draw=black, thick, text width=7cm, minimum height=0.6cm, align=center, font=\sffamily\scriptsize},
					arrow/.style={->, >=Stealth, thick, black}
					]
					\node (app) [layer, fill=white] {\textbf{Managed App (.NET 8.0)}: \textit{Classic/Diff Memento}};
					\node (clr) [layer, below=of app, fill=gray!5] {\textbf{CLR}: \textit{Server GC \& JIT Engine}};
					\node (os) [layer, below=of clr, fill=gray!15] {\textbf{OS}: \textit{Win 10 Power Mgmt}};
					\node (hw) [layer, below=of os, fill=gray!30] {\textbf{Hardware}: \textit{Intel Core / RAPL}};
					\draw[arrow] (app) -- (clr);
					\draw[arrow] (clr) -- (os);
					\draw[arrow] (os) -- (hw);
				\end{tikzpicture}%
			}
			\caption{Software-to-Hardware Stack.}
			\label{fig:architecture}
		\end{minipage}\hfill
		\begin{minipage}[t]{0.48\textwidth}
			\vspace{0pt}
			\makeatletter\def\@captype{table}\makeatother
			\centering
			\caption{Hardware \& Software Config.}
			\label{tab:specs}
			\renewcommand{\arraystretch}{0.9}
			\scriptsize
			\begin{tabular}{@{}ll@{}}
				\toprule
				\textbf{Component} & \textbf{Specification} \\ \midrule
				\textbf{CPU} & Intel i7-10700K (8C/16T) \\
				\textbf{Freq.} & Locked @ 2.9 GHz \\
				\textbf{RAM/Cache} & 32 GB DDR4 / 16 MB L3 \\
				\textbf{Runtime} & .NET 8.0, Server GC \\
				\textbf{Telemetry} & RAPL (MSR 0x610/0x611) \\ \bottomrule
			\end{tabular}
		\end{minipage}
	\end{figure}
	\FloatBarrier
	\textbf{Implementation and Control:} All benchmarks executed within a deterministic .NET 8.0 (Server GC) environment across three variants: \textbf{Baseline} (direct persistence), \textbf{Classic} (full-state serialization), and \textbf{Differential} (XOR delta-encoding). To suppress OS-induced variability, we implemented strict ``Green Mining'' protocols \cite{Hindle2012}:
	(1) CPU frequency was hard-locked at 2.9 GHz with Turbo Boost disabled;
	(2) threads were pinned to a dedicated physical core;
	(3) a 5-second thermal cool-down was enforced.
	We executed 30 discrete iterations per state size (10 MB to 200 MB), re-executing anomalies exhibiting a CV $> 5\%$.
	\section{Experimental Results and Analysis}
	\label{sec:results}
	\subsection{Energy Model Formalization}
	Total system energy ($E_{Total}$) bifurcates into two regimes. The \textbf{I/O-Bound Regime (Classic)} follows a linear model relative to state volume ($S$):
	\begin{equation}
		E_{S_{cl}}(S) \approx
		\underbrace{\kappa_{IO} \cdot S}_{\text{I/O Cost}} +
		\underbrace{\phi_{Base}}_{\text{Static Cost}} +
		\underbrace{\epsilon}_{\text{System Noise}}
	\end{equation}
	The \textbf{CPU-Bound Regime (Differential)} is driven by traversal complexity ($D$) and mutation magnitude ($\Delta U$):
	\begin{equation}
		E_{S_{diff}}(S, \Delta U) \approx
		\gamma_{CPU} \cdot f(D) +
		\delta_{DRAM} \cdot \Delta U +
		\Omega_{GC}
	\end{equation}
	where $\Omega_{GC}$ isolates the non-linear energy penalty of garbage collection.
	\subsection{The Sweet Spot vs. The Memory Wall}
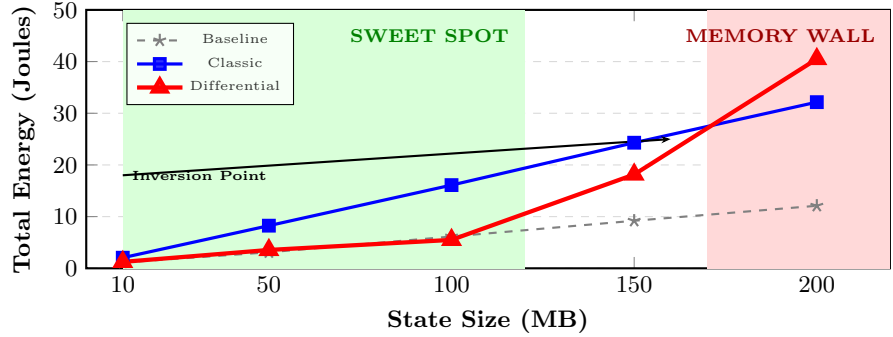
\begin{figure}[htbp]
		\centering
		\begin{tikzpicture}
			\begin{axis}[
				width=\linewidth,
				height=5cm,
				xlabel={\textbf{State Size (MB)}},
				ylabel={\textbf{Total Energy (Joules)}},
				xmin=0, xmax=220, ymin=0, ymax=50,
				xtick={10, 50, 100, 150, 200},
				ytick={0, 10, 20, 30, 40, 50},
				legend style={at={(0.05,0.95)}, anchor=north west, font=\tiny, fill=white, fill opacity=0.8},
				ymajorgrids=true,
				grid style={dashed, gray!30},
				axis line style={thick}
				]
				\fill[green!15] (axis cs:10,0) rectangle (axis cs:120,50);
				\node[green!50!black, font=\scriptsize\bfseries] at (axis cs:94, 45) {SWEET SPOT};
				\fill[red!15] (axis cs:170,0) rectangle (axis cs:220,50);
				\node[red!60!black, font=\scriptsize\bfseries] at (axis cs:190, 45) {MEMORY WALL};
				\addplot[gray, dashed, mark=star, mark size=2.5pt, line width=0.8pt]
				coordinates {(10, 1.12)(50, 3.1)(100, 6.15)(150, 9.2)(200, 12.1)};
				\addlegendentry{Baseline}
				\addplot[blue, mark=square*, mark size=2pt, line width=1.2pt]
				coordinates {(10, 2.05)(50, 8.25)(100, 16.1)(150, 24.3)(200, 32.15)};
				\addlegendentry{Classic}
				\addplot[red, ultra thick, mark=triangle*, mark size=3pt]
				coordinates {(10, 1.24)(50, 3.55)(100, 5.5)(150, 18.15)(200, 40.5)};
				\addlegendentry{Differential}
				\draw[thick, <-, >=Stealth] (axis cs:160, 25) -- (axis cs:10, 18)
				node[anchor=west, font=\tiny\bfseries, align=left] {Inversion Point};
			\end{axis}
		\end{tikzpicture}
		\vspace{0.2mm}
		\caption{Energy Dynamics and Efficiency Inversion. The green zone identifies the Sweet Spot where Differential logic succeeds, while the red zone marks the architectural collapse.}
		\label{fig:energy_dynamics}
	\end{figure}

	The aggregated data (Table~\ref{tab:energy_data}, Fig.~\ref{fig:energy_dynamics}) exposes a dramatic inflection point in efficiency.
	\begin{table}[!ht]
		\centering
		\footnotesize
		\caption{Energy (J) and GC Telemetry by State Size ($N=30$, $\alpha=0.05$)}
		\label{tab:energy_data}
		\begin{tabular*}{\linewidth}{@{\extracolsep{\fill}}lccccc}
			\toprule
			\textbf{Size} & \textbf{Baseline} & \textbf{Classic} & \textbf{Differential} & \textbf{Energy Delta} & \textbf{Gen2 GC Impact} \\ \midrule
			10 MB & 1.12 J & 2.05 J & 1.24 J & -39.5\% & Minimal \\
			100 MB & 6.15 J & 16.10 J & 5.50 J & \textbf{-65.8\%} & Stable \\
			150 MB & 9.20 J & 24.30 J & 18.15 J & -25.3\% & Escalating \\
			200 MB & 12.10 J & 32.15 J & 40.50 J & \textcolor{red}{\textbf{+25.9\%}} & \textbf{+2300\% Spike} \\ \bottomrule
		\end{tabular*}
	\end{table}
	From 10\,MB to 100\,MB, the Differential strategy maintains absolute superiority, achieving a peak efficiency delta of \textbf{-65.8\%} by aggressively suppressing DRAM activation. However, beyond the 150\,MB threshold, an exponential escalation in GC activity triggers a catastrophic collapse. At 200\,MB, the workload causes a 2,300\% surge in Generation 2 collections for the Differential approach. Its energy footprint surges to 40.5 Joules, eclipsing the na\"ive Classic model.
	\FloatBarrier
	\section{Discussion and Architectural Implications}
	The empirical evidence challenges a pervasive heuristic: the assumption that minimizing data payload inherently translates to superior system-level energy efficiency.
	\subsection{The Managed Runtime Paradox}
	We formalize a paradox where minimizing storage footprint inflates energy demand. Differential logic reduces data size but triggers hyper-allocation of ephemeral objects. In generational heaps, this velocity accelerates GC pressure, causing premature object promotion and costly Mark--Sweep--Compact cycles. This metabolic overhead explains the energy surge beyond 150~MB, where runtime management costs eclipse algorithmic gains.
	\subsection{Decision Framework}
	Rejecting a ``silver bullet'' approach, we propose an evidence-based decision matrix (Table~\ref{tab:decision_matrix}). Energy-aware architectures must dynamically pivot state-management logic based on real-time memory pressure regimes.
	\begin{table}[htbp]
		\caption{Green Architect Decision Matrix for State Persistence}
		\label{tab:decision_matrix}
		\centering
		\footnotesize
		\setlength{\tabcolsep}{4pt}
		\begin{tabular}{@{}p{3cm} p{2cm} p{7cm}@{}}
			\toprule
			\textbf{State Regime} & \textbf{Pattern} & \textbf{Architectural Rationale} \\ \midrule
			\textbf{Low Pressure} \newline ($<$ 100 MB) & \textbf{Differential} \newline ($S_{diff}$) & \textit{CPU-Bound Efficiency.} \newline Max ${\sim}65\%$ savings via suppressed DRAM traffic. \\ \midrule
			\textbf{Transition} \newline (100--150 MB) & \textbf{Hybrid/ Adaptive} & \textit{Unstable Zone.} \newline Real-time telemetry required; pivot if Gen1 GC spikes. \\ \midrule
			\textbf{Saturation} \newline ($>$ 150 MB) & \textbf{Classic} \newline ($S_{cl}$) & \textit{Memory-Bound Stability.} \newline Bypasses non-linear GC penalties. \\ \bottomrule
		\end{tabular}
	\end{table}
	\noindent Contextually, while RAPL guarantees high-resolution hardware telemetry, the exact numerical threshold of the Memory Wall is context-dependent. Conducted on an Intel CISC architecture with .NET Server GC, the inversion point may shift under alternative topographies (e.g., ARM) or concurrent memory collectors (e.g., Java ZGC).
	\section{Conclusion}
	This research characterizes the energy footprint of the Memento pattern, challenging the assumption that data volume reduction is a universal proxy for sustainability. Our findings reveal a critical paradox: aggressive delta-encoding strategies trigger catastrophic energy penalties beyond specific scaling thresholds in managed runtimes.
	We formalize this as the \textbf{``Memory Wall of Green Software''}---the point where the metabolic weight of runtime overhead (Gen2 GC and heap compaction) eclipses algorithmic gains. While Differential strategies optimize mid-scale workloads, their efficiency is non-monotonic and governed by object lifecycle dynamics. Consequently, \textbf{scalability must be treated as a fundamental energy metric}. Evaluating sustainable architecture requires a hardware-software interface perspective, acknowledging the non-linear ``energy tax'' of abstractions. This study establishes the groundwork for ``Runtime-Aware Architecture,'' ensuring future systems are sustainable by design.
	
\end{document}